\documentclass[aps,prl,twocolumn,showpacs,superscriptaddress,groupedaddress]{revtex4}  
\usepackage{graphicx}
\usepackage{dcolumn}
\usepackage{bm}
\usepackage{amssymb}
\usepackage{amsmath}
\usepackage{subfigure}
\usepackage{upgreek}

\def\registered{{\ooalign{\hfil\raise .00ex\hbox{\scriptsizeR}\hfil\crcr\mathhexbox20D}}}

\begin{document}

\title{Dynamics of self-propelled Janus particles in viscoelastic fluids}
\date{\today}

\author{Juan Ruben Gomez-Solano}
\affiliation{2. Physikalisches Institut, Universit\"at Stuttgart, Pfaffenwaldring 57, 70569 Stuttgart, Germany}
\author{Alex Blokhuis}
\affiliation{2. Physikalisches Institut, Universit\"at Stuttgart, Pfaffenwaldring 57, 70569 Stuttgart, Germany}
\author{Clemens Bechinger}
\affiliation{2. Physikalisches Institut, Universit\"at Stuttgart, Pfaffenwaldring 57, 70569 Stuttgart, Germany}
\affiliation{Max-Planck-Institute for Intelligent Systems, Heisenbergstrasse 3, 70569 Stuttgart, Germany}

\begin{abstract}

We experimentally investigate active motion of spherical Janus colloidal particles in a viscoelastic fluid. Self propulsion is achieved by a local  concentration gradient of a critical polymer mixture  which is imposed by laser illumination. Even in the regime where the fluid's viscosity is independent from the deformation rate induced by the particle, we find a remarkable increase of up to two orders of magnitude of the rotational diffusion with increasing particle velocity, which can be phenomenologically described by an effective rotational diffusion coefficient dependent on the Weissenberg number. We show that this effect gives rise to a highly anisotropic response of microswimmers in viscoelastic media to external forces depending on its orientation.
\end{abstract}


\maketitle

Nature offers a plethora of microswimmers moving in complex fluid environments~\cite{elgeti}, whose properties can deviate from Newtonian behavior due to the presence of suspended macromolecules and colloidal particles~\cite{lauga}. Some examples are bacteria in polymeric solutions~\cite{houry}, spermatozoa in cervical mucus~\cite{fauci} and microbial pathogens in stomach mucus~\cite{celli}. All these fluids are viscoelastic, i.e. they may exhibit either liquid- or solid-like behavior depending on imposed deformation rates.
Understanding the dynamics of such kind of  microscopic sytems is a topic of fundamental significance in statistichal mechanics, as they exhibit new types of non-equilibrium processes~\cite{romanczuk}. 

\begin{figure}
	\includegraphics[width=\columnwidth]{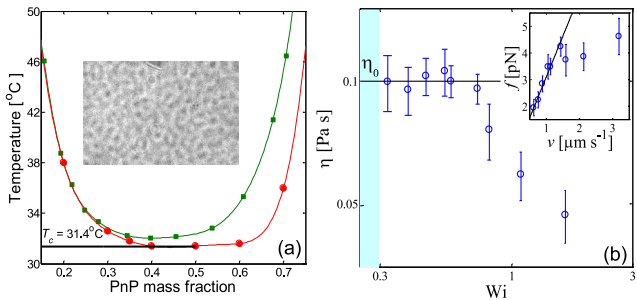}
	\caption{(color online)
(a) Phase diagram of a viscoelastic mixture of PnP and an aqueous  (0.05\% mass) PAAm solution (circles) and of the same binary mixture without polymer (squares). Inset: snapshot of the spinodal decomposition of the viscoelastic mixture at 0.4 PnP mass fraction and $32^{\circ}$C. (b) Viscosity of the viscoelastic mixture as a function of  Weissenberg number determined by active microrheology. Inset: Drag force as a function of the imposed velocity. The solid line represents the Stokes law.}
\label{fig:fig1}
\end{figure}

Despite their biological and application-related relevance, most experiments with autonomous synthetic microswimmers which are self-propelled e.g. by diffusiophoresis~\cite{howse,palacci,buttinoni} and thermophoresis~\cite{jiang}, were performed in Newtonian fluids~\cite{elgeti}. In contrast, only few studies have considered non-Newtonian fluids~\cite{teran,shen,liu,zhu,espinosa,gagnon,spagnolie,thomases,riley,qiu,martinez,datt,qin,patteson,corato} where viscoelasticity~\cite{teran,shen,liu,espinosa,gagnon,spagnolie}, shear-thinning~\cite{qiu,martinez,datt,qin} and shear thickening~\cite{qiu} strongly impact self propulsion. Previous studies with biological microswimmers suggest that, under such conditions, the dynamical response of the liquid to configurational body changes (e.g. flagellar and undulatory motion) during self propulsion must be considered~\cite{spagnolie,thomases} and can either lead to an increase~\cite{teran,liu,riley,gagnon,spagnolie,thomases,patteson}, decrease~\cite{shen,zhu,spagnolie,thomases,qin} or no change~\cite{martinez} of their swimming speed. To avoid specific effects due to such configurational changes and to focus on how the transient strain of viscoelastic fluids couples to the swimmer's stochastic motion, experiments with rigid microswimmers of simple shape are required.

In the present Letter, we study the motion of artificial microswimmers in a viscoelastic fluid. Their self propulsion is achieved by light illumination, which allows to adjust the propulsion velocity. Contrary to Newtonian liquids, we observe a drastic increase of their rotational diffusion with increasing velocity.
Such enhancement is independent on whether the velocity is acquired by self propulsion or by an external force, and can be quantitatively described by a rotational diffusion coefficient dependent on the Weissenberg number.  We find that the microstructural relaxation of the fluid due to the particle's directed motion leads to a strong coupling of rotational to translational dynamics with striking consequences for self propulsion in presence of external bias.

\begin{figure}
	\includegraphics[width=\columnwidth]{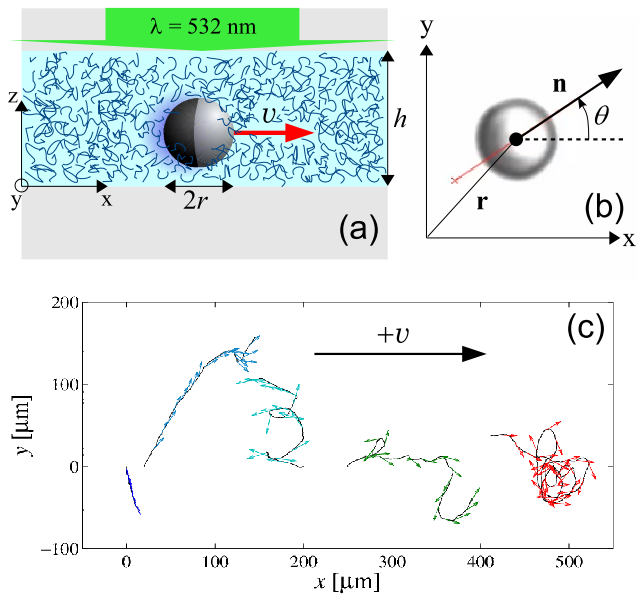}
\caption{(color online)
(a) Schematic illustration of the self-propulsion of Janus particles by light-induced demixing of the viscoelastic fluid. (b) Relevant coordinates to describe the 2D particle's motion.
(c) Examples of trajectories of the self-propelled particles at different velocities. From left to right: $v = 0.032\,\mu\mathrm{m}\,\mathrm{s}^{-1}$,  $0.188\,\mu\mathrm{m}\,\mathrm{s}^{-1}$, $0.227\,\mu\mathrm{m}\,\mathrm{s}^{-1}$, $0.398\,\mu\mathrm{m}\,\mathrm{s}^{-1}$, and $0.534\,\mu\mathrm{m}\,\mathrm{s}^{-1}$. The arrows represent the orientation $\bf{n}$.}
\label{fig:fig2}
\end{figure}

As artificial microswimmers we use half-coated spherical silica particles (diameters $2r = 7.75 \,\mu\mathrm{m}$ and $4.32 \,\mu\mathrm{m}$) coated with 50~nm Carbon caps. When such particles are suspended in a binary mixture with a lower critical point and illuminated with light, the fluid locally demixes, which causes a self-diffusiophoretic  motion (for details see~\cite{buttinoni}). As binary mixture we use water and propylene glycol propyl ether (PnP), whose lower critical point is $31.9^{\circ}$C and 0.4 PnP mass fraction. 
Such mixture exhibits Newtonian behavior with viscosity $\eta = 0.004$~Pa~s at $25^{\circ}$C. 
To render this mixture viscoelastic, we add 0.05\% polyacrylamide (PAAm). Fig.~\ref{fig:fig1} shows the corresponding phase diagrams of the binary and the ternary system and demonstrates that the addition of PAAm only leads to a small shift of the critical point. Below $T_c = 31.4^{\circ}$C the ternary mixture is homogeneous and increasing the temperature above $T_c$ leads to phase separation via spinodal decomposition, as shown in the  inset of Fig.~\ref{fig:fig1}(a).

In contrast to the critical behavior, rheological properties of the water-PnP mixture are strongly modified by the presence of PAAm polymer. This is demonstrated by active microrheology, where a colloidal particle is driven by optical tweezers through the mixture ~\cite{gomezsolano}. We find that the fluid is viscoelastic with a stress-relaxation time $\tau = 1.65\pm 0.10$~s. We obtain the fluid's viscosity $\eta$ by moving the particle at constant velocity $v$ and measuring the resulting drag force $f = 6\pi r v \eta$. The corresponding flow curve is shown in Fig.~\ref{fig:fig1}(b), where we plot the dependence of $\eta$ on the Weissenberg number, $\mathrm{Wi} = \frac{v \tau}{2r}$, where $\tau$ is the fluid's relaxation time and $v/(2r)$ the typical rate of deformation. At low Wi, the fluid is characterized by a zero-shear viscosity $\eta_0 = 0.100 \pm 0.015$~Pa~s (solid line), where the Stokes law $f \propto v$ is valid, as shown in the inset of Fig.~\ref{fig:fig1}(b). At larger Wi, we find a thinning behavior which is characteristic of most polymeric solutions, where $\eta$ decreases with increasing Wi~\cite{gomezsolano,gomezsolano1}.

\begin{figure*}
	\includegraphics[width=1.65\columnwidth]{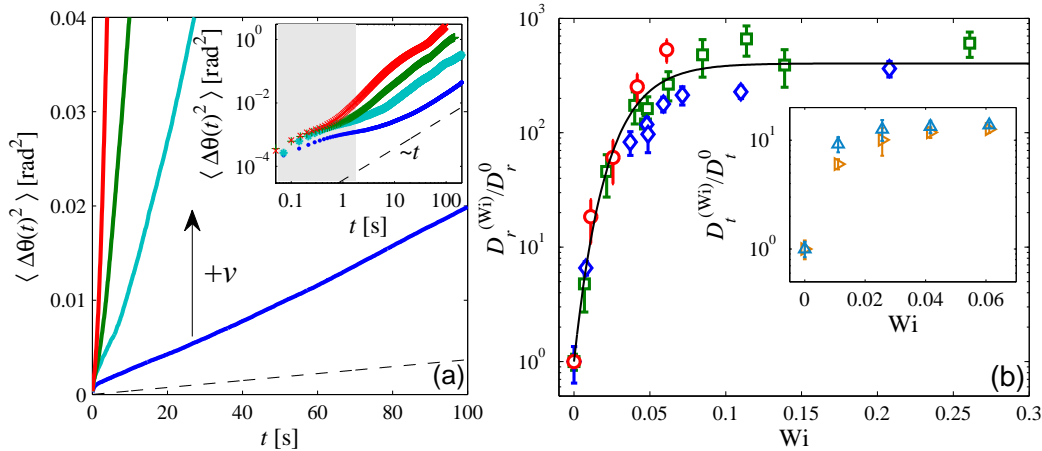}
\caption{(color online)
(a) Mean-square angular displacement for an active particle ($2r = 7.75\,\mu\mathrm{m}$) moving at different velocities $v$ in the viscoelastic fluid. From bottom to top: $v = 0.032\,\mu\mathrm{m}\,\mathrm{s}^{-1}$, $ 0.101\,\mu\mathrm{m}\,\mathrm{s}^{-1}$, $ 0.188\,\mu\mathrm{m}\,\mathrm{s}^{-1}$, $ 0.398\,\mu\mathrm{m}\,\mathrm{s}^{-1}$. The dashed line represents the case of a particle in a Newtonian fluid with the same $\eta_0$ for all $v$. Inset: log-log representation of the main figure. (b) Rotational diffusion coefficient as a function of the Weissenberg number for self-propelled particles of diameter $2r = 7.75\,\mu\mathrm{m}$ ($\Box$) and $2r = 4.32\,\mu\mathrm{m}$ ($\diamondsuit$) and a passive particle ($2r = 7.75\,\mu\mathrm{m}$)  under gravity ($\bigcirc$). Inset: Translational diffusion coefficient perpendicular ($\vartriangleright$) and parallel ($\triangle$) to the applied force for a passive particle ($2r = 7.75\,\mu\mathrm{m}$)  under gravity.}
\label{fig:fig3}
\end{figure*}

The experimental situation is represented in Fig.~\ref{fig:fig2}(a). A small volume of a dilute suspension of Janus particles in the viscoelastic mixture is confined in a cell (height $h = 4r$). Because of this confinement, both translational and rotational dynamics are limited to 2D~\cite{das}. The cell temperature is kept constant at $T = 293\,\mathrm{K} < T_c$ and then self-propulsion is induced by homogeneous laser illumination ($\lambda = 532$~nm), where the propulsion velocity is set by the light intensity ~\cite{buttinoni}. Under our conditions, particles always move opposite to the cap~\cite{samin}, as sketched in Fig.~\ref{fig:fig2}(a). Particle's position ${\bf r} = (x,y)$ and orientation ${\bf n} = (\cos\theta,\sin\theta)$ are determined from video images by detecting its barycenter and the contrast between its capped and uncapped sides, respectively (Fig.~\ref{fig:fig2}(b)). At a fixed laser intensity, we determine the mean particle velocity $v$ from the time evolution of ${\bf r}$, which allows us to compute Wi.
In all the experiments shown below, $v < 1\,\mu\mathrm{m}\,\mathrm{s}^{-1}$, which correspond to $\mathrm{Wi} < 0.3$. Accordingly, all our measurements were performed in the regime where $\eta$ is independent of Wi and equal to $\eta_0$ (cyan-shaded region in Fig~\ref{fig:fig1}(b)). In absence of laser illumination, the translational and rotational diffusion coefficients determined from experiments are $D^0_t = (3.69 \pm  0.60)\times 10^{-4}\,\mu \mathrm{m}^2\,\mathrm{s}^{-1}$ and $D^0_r = (1.84 \pm  0.40)\times 10^{-5}\,\mathrm{s}^{-1}$ for a $2r = 7.75\,\mu\mathrm{m}$ particle, respectively, which corresponds to $\eta_0 = 0.150 \pm 0.025$~Pa~s.
This value of $\eta_0$ will be used in the remainder of the paper~\footnote{This value is somewhat above the corresponding viscosity measured from active microrheology because the latter was determined in three-dimensional samples, whereas the motion of Janus particles is subject to hydrodynamic coupling to the walls.}.

Fig.~\ref{fig:fig2}(c) shows typical trajectories of a self-propelled particle measured over $1500$~seconds for increasing (left to right) laser intensities. The arrows show the corresponding particle orientation $\bf{n}$. In contrast to Newtonian liquids, where rotational diffusion is constant and independent of the propulsion velocity, here this is obviously not the case as qualitatively seen by increasing curvatures of the trajectories with increasing $v$. While at small $v$ the trajectories are rather straight, at high $v$ the particle performs several turns over the same time interval. For comparision, the rotational diffusion time of our particles in a Newtonian liquid with $\eta_0 = 0.150$~Pa~s is $1/D^0_r = 15$~hours, i.e. the particle would not show a visible change in orientation during $1500$~seconds.

To quantify the above observations, we compute the mean-square displacement of the angular particle orientation $\theta$,
$\langle \Delta \theta (t)^2 \rangle = \langle [\theta(t_0 + t) - \theta(t_0)]^2  \rangle $, where the brackets denote an average over the time variable $t_0$.
The results are shown for different velocities $v$ in Fig.~\ref{fig:fig3}(a). The subdiffusive behavior at $t \lesssim \tau$ (shaded area in inset of Fig.~~\ref{fig:fig3}(a)) is a result of the elasticity of the fluid~\cite{cheng,andablo,haro}. For $t \gtrsim \tau$, $\langle \Delta \theta (t)^2 \rangle $ shows a diffusive behavior. However, the slope of $\langle \Delta \theta (t)^2 \rangle $ strongly deviates from $2D^0_r$, i.e. from the value 
expected for a Newtonian fluid (dashed line in Fig.~\ref{fig:fig3}(a)) and for passive colloidal probes in thermal equilibrium with the surrounding viscoelastic fluid~\cite{colin}. Instead, the slope  increases with increasing $v$. 
Phenomenologically, this corresponds to an effective rotational diffusion coefficient $D^{(\mathrm{Wi})}_r$, which depends on the Weissenberg number
\begin{equation}\label{eq:effDr}
	\langle \Delta \theta (t)^2 \rangle  = 2D^{(\mathrm{Wi})}_r t.
\end{equation}
In Fig.~\ref{fig:fig3}(b) we plot as squares the dependence of $D^{(\mathrm{Wi})}_r$, normalized by $D^0_r$, as a function of Wi for a $2r = 7.75\,\mu\mathrm{m}$ particle. We find that $D^{(\mathrm{Wi})}_r$ dramatically increases and saturates above Wi$\approx 0.1$ at a value which is about $400 D^0_r$. An identical behavior is observed also for other particle sizes (diamonds correspond to particles with $4.23\,\mu\mathrm{m}$ in diameter) and thus hints at a size-independent feature.

The previous findings suggest that such a diffusive enhancement is a generic feature originating from the flow field around the particle. In fact, unlike in a Newtonian liquid, a non-steady flow is induced as the particle moves through a viscoelastic fluid, since the relaxation of the local stress takes place over a time-scale $\approx \tau$, during which the fluid is driven out of equilibrium. Consequently, such a fluctuating flow field can exert random forces and torques on the particle coupled to those originating from thermal collisions with the fluid molecules, which can be interpreted as enhanced diffusion. A similar effect has been observed in active microrheology of dense colloidal suspensions~\cite{wilson,winter}, where the coupling between a driven particle and the slow structural relaxation of the suspension enhances the non-equilibrium fluctuations of the particle position.
Thus, an enhancement of of both rotational and translational diffusion should also be observed when driving a particle through a viscoelastic fluid by an external field. 
Therefore, we also measure the rotational diffusion coefficient $D^{(\mathrm{Wi})}_r$ for the same Janus particles while sedimenting in a tilted sample cell, without laser illumination to avoid self propulsion~\footnote{The rotational motion of non-activated particles takes place in 3D. Nevertheless, $D^{(\mathrm{Wi})}_r$ can still be determined from Eq.~(\ref{eq:effDr}), because $\theta$ results from the 2D projection of \textbf{n} on the x-y plane.}. At a inclination angle $\alpha$, the particle is dragged through the fluid by a gravitational force ${\bf F}_g = {\bf g} m_b \sin \alpha$, where ${\bf g}$ is the acceleration of gravity and  $m_b$ the particle's effective mass. In Fig.~\ref{fig:fig3}(b) we plot as circles $D^{(\mathrm{Wi})}_r / D^0_r$ as a function of Wi (obtained from the mean settling velocity $v$) for a $2r = 7.75\,\mu\mathrm{m}$ particle. Although self propulsion and passive sedimentation of Janus particles display entirely different translational dynamics,
we find an almost identical behavior of $D^{(\mathrm{Wi})}_r / D^0_r$.
This quantitative agreement is at first glance surprising because the flow field and thus the strain in the liquid is not generally identical in both cases~\cite{drescher}. Hence, we provide evidence that the flow field generated by the Janus microswimmers in our experiments is Stokesian-like, as recently proposed by~\cite{samin}.
Furthermore, we observe that the translational diffusion coefficient $D^{(\mathrm{Wi})}_t$ is also enhanced with increasing Wi, as shown in the inset of Fig.~\ref{fig:fig3}(b). However, the enhancement is not as pronounced as the rotational one. This is a common feature of non-equilibrium fluctuations, whose properties are frequently observable-dependent and where an effective temperature is well-defined only under very specific conditions~\cite{seifert,dieterich}.

\begin{figure}
	\includegraphics[width=2.8in]{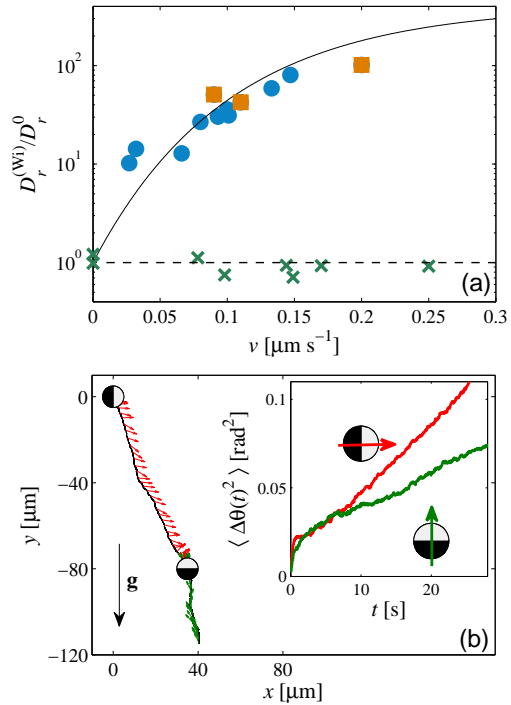}
\caption{(color online)
(a) Rotational diffusion coefficient of a $2r = 7.75\,\mu\mathrm{m}$ particle as a function of its velocity resulting from different laser intensities and distinct inclination angles, moving in the viscoelastic fluid at  $\alpha = 5^{\circ}$ ($\bigcirc$), $\alpha = 25^{\circ}$ ($\Box$) and in a Newtonian fluid at $\alpha = 5^{\circ}$ ($\times$). Solid line: same guide to the eye as in Fig.~\ref{fig:fig4}(b), dashed line: Stokes-Einstein relation. (b) Trajectory and orientation of a self-propelled particle moving in  the viscoelastic fluid, with an average initial orientation perpendicular (red arrows) and then antiparallel (green arrows) to ${\bf{g}}$. Inset: mean-square angular displacement corresponding to these two configurations.}
\label{fig:fig4}
\end{figure}

Our results show that the enhancement of rotational diffusion is independent on whether the particle velocity is caused
by self propulsion or by an external field. Therefore, the same effective description must be valid for a self-propelled particle under gravity. In this case, the relevant parameter for the Weissenberg number is the total velocity $v = \left| {\bf v}_{sp} + \frac{{\bf F}_g}{6\pi\eta_0 r} \right|$,
resulting from the propulsion velocity $ {\bf v}_{sp}$ and the settling velocity $\frac{{\bf F}_g}{6\pi\eta_0 r}$. 
We demonstrate this in Fig.~\ref{fig:fig4}(a), where we plot as circles and squares the dependence of $D^{(\mathrm{Wi})}_r$ on $v$ for $2r = 7.75\,\mu\mathrm{m}$ Janus particles at two different inclination angles $\alpha$ and for different laser intensities. We find that the combination of self propulsion and gravity leads to an enhancement of the rotational diffusion quantitatively similar to that plotted in Fig.~\ref{fig:fig3}(b), whose phenomenological fit  is represented as solid lines both in Figs.~\ref{fig:fig3}(b) and \ref{fig:fig4}(a).

Although the above effective description of rotational diffusion of microswimmers in a viscoelastic fluid is rather simple, it has profound implications for their response to external forces~\cite{hagen,ginot}. We point out that in a Newtonian liquid, rotational diffusion of a colloidal particle originates solely from the collisions with the fluid molecules, thus resulting in a complete  decoupling of rotational from translational motion regardless of whether the particle is passive, self-propelled, or externally driven. Indeed, in Fig.~\ref{fig:fig4}(a) we show as crosses the rotational diffusion coefficient of a Janus particle ($2r = 7.75\,\mu\mathrm{m}$) actively moving though a Newtonian liquid (water-PnP) under gravity ($\alpha = 5^{\circ}$) at different laser intensities. In this case, even for similar velocities as those achieved in the viscoelastic mixture, the rotational diffusion coefficient is independent from $v$ and equal to that given by the Stokes-Einstein relation $D^0_r = k_B T /(8\pi\eta_0 r^3)$. 
This is in stark contrast to viscoelastic fluids, where the coupling between translational and orientational degrees of freedom is expected to give rise to a strong dynamical dependence on the particle orientation in presence of an external force. In Fig.~\ref{fig:fig4}(b) we demonstrate this effect, where an active particle, whose velocity at $\alpha = 0^{\circ}$ is $v_{sp} = 0.080\mu\mathrm{m}\,\mathrm{s}^{-1}$,  is subject to gravity ($\alpha = 25^{\circ}$), at which the settling velocity is  $0.180\mu\mathrm{m}\,\mathrm{s}^{-1}$. Depending on its orientation relative to gravity $\bf{g}$, different velocities $v$ can be achieved. First, when the orientation is approximately perpendicular to $\bf g$ ($\theta \approx 0^{\circ}$), the mean velocity ($v \approx 0.200\mu\mathrm{m}\,\mathrm{s}^{-1}$) leads to the mean-square angular displacement plotted in the inset of Fig.~\ref{fig:fig4}(b). Its effective rotational diffusion coefficient with this orientation is $D^{(\mathrm{Wi})}_r \approx 102 D^{0}_r$.
Due to such rather large value of $D^{(\mathrm{Wi})}_r$, the particle is able to perform a rotation of $90^{\circ}$ during the measurement time (10 minutes).
With its new orientation ($\theta \approx 90^{\circ}$), i.e. antiparallel to {\bf g}, the resulting mean velocity is smaller ($v \approx 0.090\mu\mathrm{m}\,\mathrm{s}^{-1}$). Then, the corresponding mean-square angular displacement has a smaller slope, as shown in the inset of Fig.~\ref{fig:fig4}(b), which corresponds to $D^{(\mathrm{Wi})}_r \approx 53 D^{0}_r$. Consequently, the probability for the particle to remain with such orientation must be higher than before, as experimentally verified. 

It should be mentioned that such a behavior is not expected to be particular to external forces, but also to other types of bias commonly found in most situations, e.g. confined geometries~\cite{das,figueroa}, external flows~\cite{zoettl,palacci1,mathijssen}, and extended gradients~\cite{jileki}. In all these cases, the microswimmer's velocity strongly depends on its position or orientation relative to, e.g. a solid wall, the flow direction, or a chemical gradient, which can result in an anisotropic response when moving in viscoelastic media.

In conclusion, we have investigated self propulsion of Janus particles in a viscoelastic fluid. Our study represents the first experimental realization of autonomous synthetic microswimmers in a non-Newtonian environment, which have allowed us to uncover the role of viscoelasticity in active Brownian motion. We have found a dramatic enhancement of the rotational diffusion of the particles with increasing propulsion velocity, which can be phenomenologically described by an effective rotational diffusion coefficient dependent on the Weissenberg number. This non-equilibrium phenomenon, absent in Newtonian fluids, arises from the coupling between the particle's directed motion and the microstructural relaxation of the surrounding fluid. We have demonstrated that this effect leads to dramatic changes in the dynamical response of self-propelled particles depending on their orientation relative to an external force. Thus, our findings have important consequences for the behavior of microswimmers subject to other types of bias, as those commonly induced by external flows, gradients, and confined geometries. 

We would like to acknowledge Mahsa Sahebdivani for her assistance in the sample preparation, Hans-J\"urgen K\"ummerer for the particle tracking analysis, and Celia Lozano for helpful discussions. This work was financially supported by the Deutsche Forschungsgemeinschaft within the Schwerpunktsprogramm Microswimmers SPP 1726.

\end{document}